# Switching of the Magnetic Vortex Core in a Pac-man Disk using a Single Field Pulse

Tomonori Sato, Keisuke Yamada, Yoshinobu Nakatani

Graduate School of Informatics and Engineering,
University of Electro-Communications, Chofu, Tokyo, 182-8585, Japan

**Abstract**

We report on the switching of the magnetic vortex core in a Pac-man disk using a magnetic field pulse, investigated via micromagnetic simulations. The minimum core switching field is reduced by 72 % compared to that of a circular disk with the same diameter and thickness. However, the core switches irregularly with respect to both the field pulse amplitude and duration. This irregularity is induced by magnetization oscillations which arise due to excitation of the spin waves when the core annihilates. We show that the core switching can be controlled with the assist magnetic field and by changing the waveform.

## Introduction

The magnetic vortex is a flux closure domain structure in ferromagnetic elements with a sub-micrometer length scale[1-10]. The magnetic vortex structure has two characteristics of a magnetic domain structure: the direction of the in-plane rotating magnetization (chirality), and the direction of the perpendicular magnetization in the center of the disk (polarity), which is called the vortex core[1,2]. Because the vortex core is very stable, a DC magnetic field of more than several thousand Oersted is required to switch the polarity[3]. Recently, various core switching techniques using the magnetic field or the spin current have been proposed[7-11]. The results achieved with these techniques may potentially lead to the construction of devices such as nonvolatile data storage. In these methods, the core is switched by creating and annihilating a vortex-antivortex (V-AV) pair[7-10]. To develop a new switching method which does not use a V-AV pair, we focus on the Pac-man (PM) disk[12,13] as a candidate for reducing the magnetic field required for core switching. The PM shape has an open slot extending from the center to the circumference. In early investigations using the PM disk, some researchers have observed the magnetic domain structure produced by changing the notch angle and measured the core switching field with an in-plane DC magnetic field[12]. Other authors have measured domain structures of both chirality and polarity by changing the thickness and the diameter of the PM disk using an in-plane DC magnetic field [13]. In this letter, we performed micromagnetic simulations and investigated the vortex core switching using an in-plane nanosecond field pulse in order to study the switching mechanisms and switching conditions in the PM disk in detail. We also investigated the effects of the assist magnetic field, the field pulse waveform, and the notch depth and angle on the core switching.

## Simulation

A three-dimensional micromagnetic model was used in the simulation; Figure 1(a) shows the PM disk which was used. The notch depth from the circumference and the notch angle were defined as $d$ and $\theta$, respectively. A field pulse was applied in the x-axis direction. The diameter ($D$) and the thickness ($h$) of the PM disk were 200 nm and 40 nm, respectively. Typical material parameters for Permalloy were used, as follows: exchange stiffness constant $A = 1\times10^{-6}$ erg/cm, saturation magnetization $M_s$ = 800 emu/cm$^3$, magnetocrystalline anisotropy $K_u$ = 0, and Gilbert damping constant α = 0.01. The PM disk was divided by a rectangular prism with dimensions of 2 ×2×2.5 nm³. The lateral and the vertical dimensions (2.0 and 2.5 nm) are safely below the exchange length of Permalloy (~5 nm). The simulations capture effects from the vortex fine structure inside the bulk material, such as breathing or meandering, which could play an important role in the switching process[4,15].

Results

Figures 1(c) to (f) show snapshots of the core switching of the PM disk using a field pulse with an amplitude of $H_x$ = 100 Oe, a duration of $t_p$ = 2.0 ns, and rise and fall times of 0 ns (Fig. 1 (b)). In the remanent state, the direction of the vortex core magnetization is upward (white color in Fig. 1 (c)). The core moves to the notch with the gyrotropic mode of the field pulse[15] (Fig. 1 (d)), and it annihilates at the notch edge (Fig. 1 (e)). After the field pulse is cut off, the downward core (black color) nucleates from the bottom part of the notch (Fig. 1 (f)), and moves to the disk center with the gyrotropic mode.

Figure 2(a) presents a diagram of the core switching in the PM disk as a function of the amplitude and duration of the field pulse. Here, the pulse duration step is 0.01 ns. The core switches in the red and blue color regions of the diagram; in contrast, it does not switch in the white regions. The annihilation mechanism of the core is different between the blue and the red regions. In the blue regions ($H_x$ < 470 Oe), it annihilates with the same mechanisms as Figs. 1 (c-e), whereas in the red regions ($H_x$ > 470 Oe), the V-AV pair appears and annihilates before the core reaches the notch edge.

In Fig. 2(a), the core switches irregularly with respect to both the field pulse amplitude and duration. In order to clarify the reason for this, we investigated the time evolution of the averaged magnetization in the perpendicular direction $m_z$ and its time derivative $d(<m_z>)/dt$ at the notch edge region before and after the field pulse is cut off (Figs. 2(b) and (c)). Here, the notch edge region is the red dotted square region in Fig. 1 (a) (12×12 nm²). The size of this region is almost same as the vortex core size. The core switches (downward core appears) in Fig. 2(b) ($t_p$ = 6.00 ns) and it does not switch (upward core appears) in Fig. 2(c) ($t_p$ = 5.90 ns). In Fig. 2(b) (Fig. 2(c)), $<m_z>$ and $d(<m_z>)/dt$ have small oscillations before the field pulse is cut off, whereas $d(<m_z>)/dt$ decreases (increases) markedly about 0.05 ns after $t_p$, then <$m_z$> decreases (increases). This time lapse of 0.05 ns is almost same in all cases. The direction of the nucleated core is same as the sign of <$m_z$> at 0.05 ns after the field pulse. A negative (positive) <$m_z$> creates a downward (upward) core, meaning that the core switches (does not switch). We define this time lapse as the nucleation time.

In order to investigate the relation between <$m_z$> and the direction of the nucleated core in more detail, we compared the time evolution of <$m_z$> (the black line) and the core switching regions up to $t_p$ =6.00 ns with $H_x$ = 100 Oe (Fig. 2(d)). The blue areas show the core switching regions, and the red area shows the dispersion of $m_z$ in the red dotted square region in Fig. 1(a). Here, the time of <$m_z$> and the dispersion are shifted by -0.05 ns (the nucleation time). Spin waves appear when the core annihilates ($t$ = 0.84 ns), then <$m_z$> oscillates intensively. The oscillations are gradually damped with time. Figure 2(e) shows the time evolution of <$m_z$> and the core switching regions between 2 and 3 ns (the red dotted square region in Fig. 2(d)). The core switching regions (blue) and non-switching regions (white) almost agree with the black line in the negative and positive <$m_z$> regions in Figs. 2(d, e). The shifts between the line of <$m_z$> and the core switching (non-switching) regions can be explained by considering the dispersion of $m_z$. These results show that the sign of <$m_z$> at 0.05 ns after the field pulse determines the direction of nucleated core, and the irregularity of the core switching in Fig. 2(a) is an effect of the spin waves generated by the core annihilation.

We compared the minimum switching field of the PM disk and the circular disk with the same diameter and thickness, which are 90 and 320 Oe, respectively. The minimum switching field is reduced by 72% in the PM disk.

The switching field decreases when using the PM disk; however, it is difficult to control the switching. We applied the assist field, which is the DC magnetic field in the perpendicular direction, and changed the waveform of the field pulse to control the switching. Figure 3(a) presents a diagram of the core switching with a square field pulse with an assist field of -50 Oe. The core switches irregularly in the short pulse case; in contrast, the core switches irrespective of the pulse duration for $t_p$ > 4.52 ns and $H_x$ > 90 Oe. We define this pulse duration as $t_p^{th}$ (here $t_p^{th}$ = 4.52 ns). Figure 3(b) shows the time evolution of the magnetization up to $t_p$ = 6.00 ns with $H_x$ = 100 Oe. Since the core nucleation time is 0.05 ns, which is the same as in Figs. 2(b, c), the times of <$m_z$> and its deviation are shifted by -0.05 ns in Fig. 3(b), as they are in Figs. 2(d, e). Although this figure is similar to Fig. 2(d), <$m_z$> has a negative value (downward) in almost all cases for $t$ > 3.16 ns. This shows that the direction of the nucleated core is the same as the direction of the assist field for $t_p$ > 3.16 ns. (We note that a positive <$m_z$> appears around $t$ ~3.84 ns. This can also be explained by considering the dispersion of $m_z$, similarly to Figs. 2(d, e)). Thus, the oscillations of <$m_z$> can be controlled by the assist field after a lapse of a predetermined amount of time, so the core switching can be controlled with the assist field for $t_p$ > $t_p^{th}$. We investigated the relation between the assist field and $t_p^{th}$ and determined the values of $t_p^{th}$ for assist field strengths of -25, -50 and -100 Oe to be 5.42, 4.52 and 3.56 ns, respectively.

Next, we changed the field pulse form to a sawtooth pulse with an assist field of -50 Oe (Fig. 3(c)). Here, $t_p^{th}$ is reduced to 2.80 ns. Figure 3(d) shows the time evolution of <$m_z$> with $H_x$ = 100 Oe and $t_p$ = 6.00 ns. The downward core nucleates before the field pulse is cut off ($t$ ~ 2.50 ns). The switching finishes at $t$ ~ 3.50 ns, since the duration time of the core switching becomes longer due to the effect of the field pulse. After the core switching, the core moves around the disk center in the gyrotropic mode, hence <$m_z$> of the notch edge increases when the core approaches the edge of the disk. Note that the center of the <$m_z$> oscillations (peak to peak) decreases in time from $t$ ~ 2.00 ns, and <$m_z$> becomes negative just before the core nucleation. This shows that the shape of the field pulse influences the time evolution of the out-of-plane magnetization (<$m_z$>), and the assist field combined with the sawtooth field pulse seems to push the out-of-plane component further down in comparison to the square field pulse. In Fig. 3(c), the core does not switch irregularly around $t_p$ ~1-2 ns. This is due to the interactions between the <$m_z$> oscillations and the reflected first spin waves from the disk edge, which are excited by the core annihilation. We also investigated the core switching with a sawtooth field pulse with a contrary slope (not shown). In this case, the core switches for $H_x$ > 140 Oe; however, it switches irregularly, similar to what is seen in Fig. 2(a), with the same field and pulse duration conditions.

Finally, we investigated the core switching mechanisms when changing the notch depth from 30 to 98 nm, stepped by 2 nm, and changing the angle from 5° to 50°, stepped by 5°, as shown in Fig. 4(a). The figure shows that the switching mechanism is less dependent on the notch angle; however, it is strongly dependent on the notch depth. The mechanism can be classified into three patterns as shown in Fig. 4(a): (1)-(3). Figs. 4(b) to (j) show snapshots of the magnetization $m_z$ (white: $m_z$ = +1, black: $m_z$ = -1). The notch angles and depths are (1)(b-d): $\theta$ = 25º, $d$ = 52 nm, (2)(e-g): $\theta$ = 25º, $d$ = 72 nm, and (3)(h-j): $\theta$ = 25º, $d$ = 92 nm. Here, we set the time to be $t$ = 0.00 ns when the field pulse is cut off. In the case where the notch depth is short (1), the core switches (Fig. 4(d)) with nucleating and annihilating the V-AV pair. Because the position of the created core is far from the disk center, the core moves at a high speed and it reaches the critical velocity ($v_c$ ~ 260 m/s)[8,9]. At the critical velocity, the V-AV is nucleated and annihilated, and finally the core switches. In the case of notch depth (2), the switching mechanism is same as Figs.1 (c-f). The nucleated core at the notch edge moves toward the center of the disk while maintaining the core polarity (Figs. 4 (e-g)). In the case where the notch depth is deep (3), the core is not created (Figs. 4(h-j)). In this shape, the single domain and vortex states correspond to two distinct minima separated by an energy barrier; as a result, the core cannot be created automatically from the single domain state. These results show that the switching mechanism strongly depends on the notch depth, and the proper notch depth should be chosen to achieve the core switching.

## Conclusions

In this letter, we have shown vortex core switching in a Pac-man disk using a nanosecond range field pulse via micromagnetic simulations. The minimum vortex core switching field of the PM disk is reduced by 72% compared to that of a circular disk with the same diameter and thickness. However, the vortex core switches irregularly with respect to both the field pulse amplitude and duration. This irregularity is induced by magnetization oscillations arising from the spin waves caused by the core annihilation. The core switching can be controlled using an assist magnetic field and a sawtooth field pulse. Moreover, we also investigated core switching mechanisms classified into three categories based on notch depth, and show that the proper notch depth should be chosen to produce the core switching.

## Acknowledgements
KY is supported by the JSPS Postdoctoral Fellowships for Research.

Figure captions

Fig. 1 (color online) (a) Simulation model of the Pac-man disk. The notch depth and notch angle are $d = 80$ nm and $\theta = 25$ degrees, respectively. The diameter and the thickness are $D = 200$ nm and $h = 40$ nm. The red dotted square shows the notch edge ($12 \times 12$ nm$^2$). (b) The waveform of the square field pulse with $H_x = 100$ Oe and $t_p = 2.00$ ns. (c-f) Snapshots of the PM disk at each time ($t$): (c) the initial state, (d) before annihilation, (e) after annihilation, (f) after nucleation. The rainbow colors indicate the in-plane component. The sense of the curl of the in-plane magnetization is counter-clockwise.

Fig. 2 (color online) (a) Diagram of the core switching as a function of the field pulse duration and amplitude in the PM disk. The pulse duration step is 0.01 ns. The core switches in the red and blue color regions, and does not switch in the white color regions. The mechanisms of the core annihilation are different between the blue and the red regions. (b, c) The time evolution of $<m_z>$ (the red line) and $d(<m_z>)/dt$ (the blue dotted line) at the notch edge (the red dotted square in Fig. 1 (a) ) around $t = 6$ ns with $H_x = 100$ Oe. (b) The core switches (downward core appears) at $t_p = 6.00$ ns. (c) The core does not switch (upward core appears) at $t_p = 5.90$ ns. (d) The time evolution of $<m_z>$ (the black line) up to $t = 6.00$ ns with $H_x = 100$ Oe. The red gradation and the blue regions are the dispersion of $m_z$ and the core switching regions. (e) The time evolution of $<m_z>$ and the core switching regions between 2 and 3 ns.

Fig. 3 (color online) Diagram of the core switching with an assist field of -50 Oe for (a) the square field pulse, and (c) the sawtooth field pulse (inset: the waveform). (b), (d) The time evolution of $<m_z>$ (the black line) at the notch edge region. The red in (b), (d) and the blue regions in (b) are the dispersion of $m_z$ and the core switching regions, respectively. The blue dotted line (d) is the waveform of the sawtooth field pulse with $H_x = 100$ Oe and $t_p = 6.00$ ns.

Fig. 4 (color online) (a) Diagram of the dependence of the switching mechanisms on the notch depth and angle. The switching mechanism in each color region is different (1)-(3). Snapshots of the magnetization, $m_z$ (white: $m_z$

= +1, black: $m_z$ = -1) for the following cases – (1)(b-d): $\theta$ = 25°, $d$ = 52 nm, (2)(e-g): $\theta$ = 25°, $d$ = 72 nm, (3)(h-j): $\theta$ = 25°, $d$ = 92 nm.

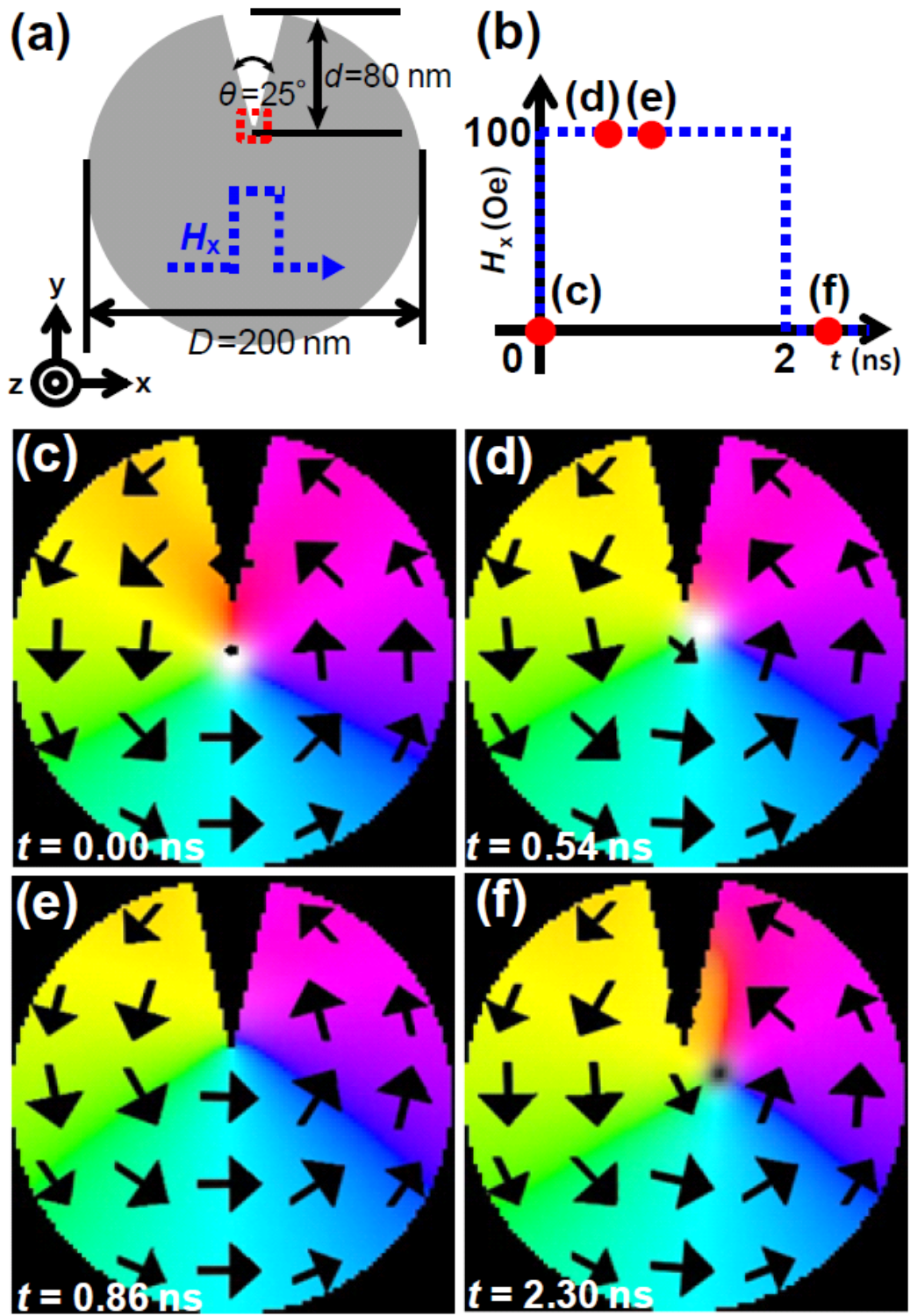

**Figure 1**

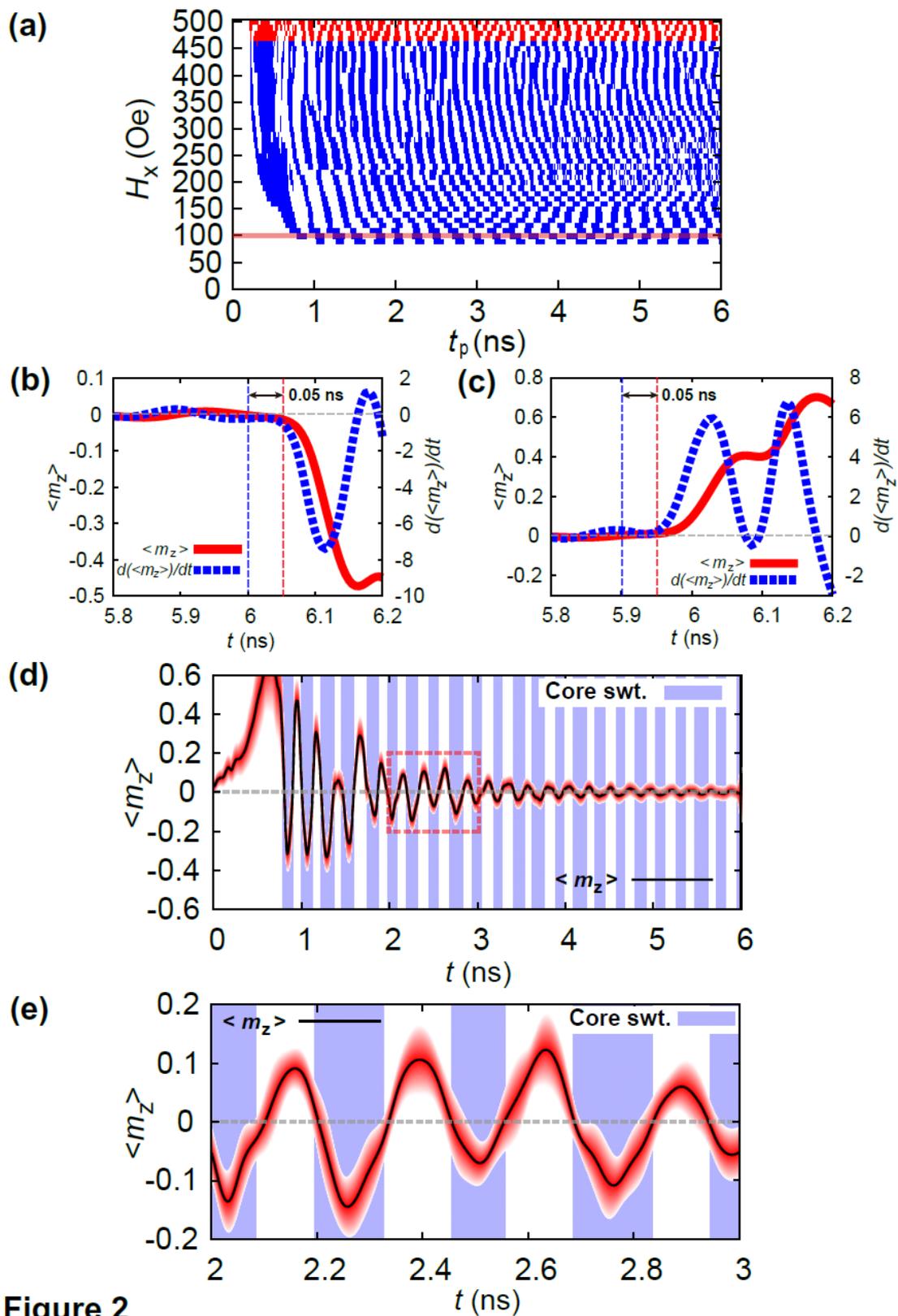

**Figure 2**

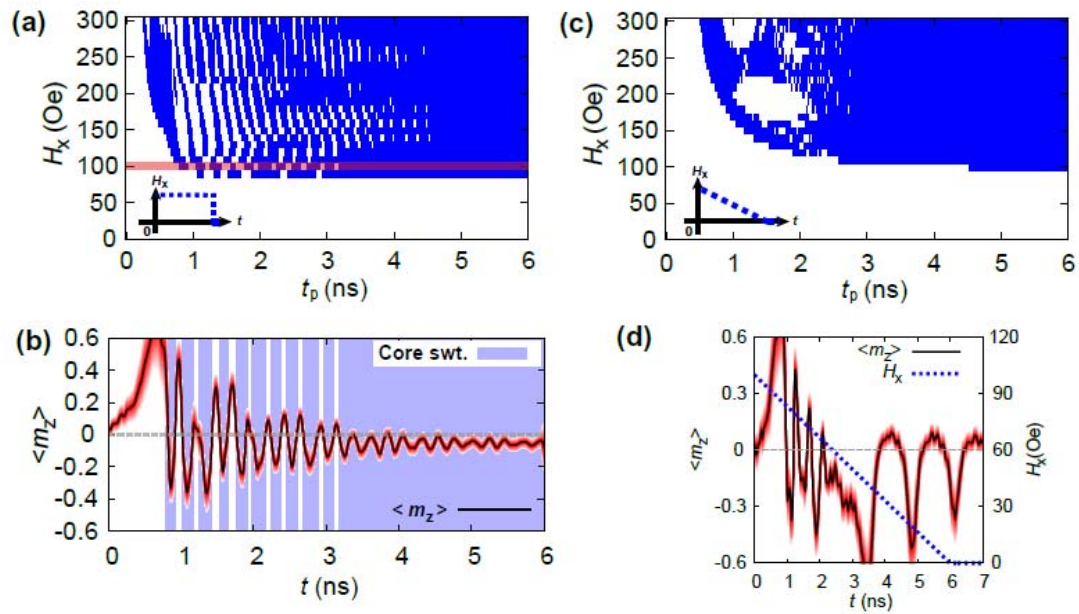

**Figure 3**

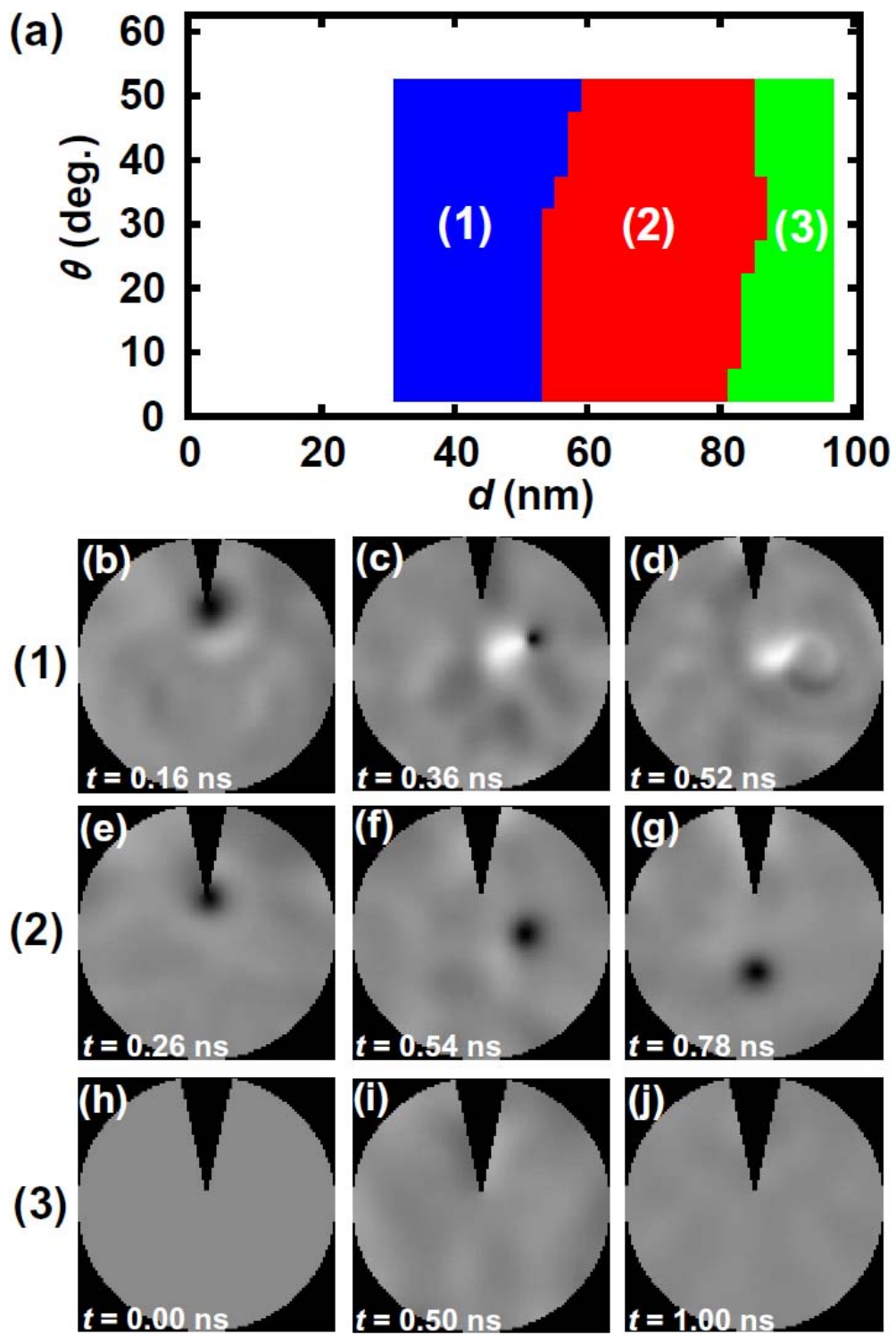

**Figure 4**